\documentclass[11pt]{article}
\usepackage[margin=1in]{geometry}
\usepackage{lmodern}
\usepackage{sidecap}
\usepackage{xspace}
\usepackage{doi}
\usepackage{hyperref} 
\usepackage{tikz}
\usepackage{pgfgantt}
\usepackage{graphicx,pbox}
\usepackage{xcolor,hyperref,colortbl}
\usepackage{multicol}
\usepackage{wrapfig}
\usepackage{siunitx}
\usepackage{xspace}
\usepackage{caption}

\def\Title#1{\begin{center} {\Large #1 } \end{center}}
\def\Author#1{\begin{center}{ \sc #1} \end{center}}

\def\Address#1{\begin{center}{ \it #1} \end{center}}

\def\CCC{C$^{3}$~}

\newcommand{\micron}{\ensuremath{\mu\mathrm{m}}}
\newcommand{\geant}{\textsc{Geant4}\xspace}
\newcommand{\SID}{{SiD}\xspace}

\newcommand\snowmass{\begin{center}\rule[-0.2in]{\hsize}{0.01in}\\\rule{\hsize}{0.01in}\\
\vskip 0.1in Submitted to the  Proceedings of the US Community Study\\ 
on the Future of Particle Physics (Snowmass 2021)\\ 
\rule{\hsize}{0.01in}\\\rule[+0.2in]{\hsize}{0.01in} \end{center}}

\usepackage[utf8]{inputenc}
\begin{document}

\snowmass

\Title{Monolithic Active Pixel Sensors on CMOS technologies}
\Author{Martin Breidenbach, Angelo Dragone, Norman Graf, Tim K. Nelson, Lorenzo Rota, Julie Segal, Christopher J. Kenney, Ryan Herbst, Gunther Haller, Thomas Markiewicz, Caterina Vernieri, Charles Young}
\Address{SLAC National Accelerator Laboratory}

\Author{Grzegorz Deptuch, Gabriella Carinii, Gabriele Giacomini, Giovanni Pinaroli, Alessandro Tricoli}
\Address{Brookhaven National Laboratory}

\Author{Nicole Apadula, Alberto Collu, Carl Grace, Leo Greiner, Yuan Mei,  Ernst Sichtermann}
\Address{Lawrence Berkeley National Laboratory}

\Author{James Brau, Nikolai Sinev, David Strom}
\Address{University of Oregon, Eugene}

\Author{Marcel Demarteau}
\Address{Oak Ridge National Laboratory}

\Author{Whitney Armstrong, Manoj Jadhav, Sylvester Joosten, Jihee Kim, Jessica Metcalfe, Zein-Eddine Meziani, Chao Peng, Paul E. Reimer, Marshall Scott, Maria \.{Z}urek}
\Address{Argonne National Laboratory}

\Author{R. Caputo, C. Kierans, A. Steinhebel}
\Address{NASA Goddard Space Flight Center}

\date{November 2021}


\tableofcontents
\newpage

\section{Introduction} 

Collider detectors have taken advantage of the resolution and accuracy of silicon (Si) detectors for at least four decades~\cite{Allport}. Future colliders will need large areas of silicon sensors, several hundred m$^{2}$, for low mass trackers and sampling calorimetry~\cite{behnke2013international}. Trackers will require multiple layers, large radii, and micron scale resolution. Sampling calorimeters will also have very large
areas and are improved by very thin overall packages, which reduces the Moliere radius. 

Monolithic Active Pixel Sensors (MAPS), in which Si diodes and readout circuitry are combined in the same pixels, and can be fabricated in some of standard CMOS processes, were shown as a promising technology for high-granularity and light in material budget detectors ~\cite{Deptuch:2002tns,Peric:2007qtk,Pohl:2017xoi}.
MAPS represent several advantages over traditional hybrid pixel detector technologies, as they can be inexpensively built, even in large array sizes, thinned to the needs, offer individual pixel readout, reasonable radiation hardness, low-mass, while operating at high speeds with low power consumption, from single and low voltage supply. The close connection of a sensor and front-end amplifier, without the need of externally added interconnections, reduces to the smallest achievable level the input capacitance. This leads to the reduction of the noise floor and thus operation with smaller magnitude signals from thinner sensors is possible with satisfactorily high S/N. 
 
Interconnects, routinely realized with some kind of metal spheres or pillars bump-bonding, are on one hand comparative in size with the most recently desired pixel sizes O(10 \micron). On the other hand, they are responsible for the dominating contribution to the capacitance seen at the input node of an in-pixel amplifier, effectively dominating otherwise very small sensor capacitance. Reduction of this capacitance, even thanks to the low-temperature direct bonding technology was shown in comparative studies~\cite{Deptuch:2016ted} to improve performance of a pixel detector significantly.  
 
Magnitude, speed and efficiency of collecting charge carriers, liberated in interactions of radiation with the medium of a sensor can be completely controlled in hybrid pixel detectors. Whilst MAPS are dependent on the substrate used in the CMOS process in terms of how thick the active sensor volume can be and whether a sensor can be operated depleted. Retaining the advantage of using standard CMOS processes for building MAPS, the older-generation MAPS, such as those used in the Heavy Flavor Tracker (HFT) at STAR ~\cite{greiner:2011nim}, were built in fabrication processes for which foundries were using low-resistivity (about 10 $\Omega$cm), epitaxial wafers. At that time, thermal diffusion charge collection allowed magnitudes of signals on the order of 1k e$^-$, shared among the neighbouring pixels and collected in times on the order of 100 ns~\cite{Deptuch:2001nim}. The significant limitations of the early MAPS was use of only one type of transistor in a pixel circuitry. The radiation hardness was on the order 10$^{13}$ of 1 MeV neutrons equivalent per cm$^2$ and 10 Mrads of the Total Ionizing Dose (TID)~\cite{Deveaux:2007nim}. This parameters were suitable for the HFT at STAR, that was the first practical use of MAPS.  MAPS, that were simultaneously developed using processes suitable for High Voltage applications, have not been used in any experiment yet.   

Another key feature of MAPS, that are fabricated in any standard CMOS process, is the limitation of a single chip dimensions to a reticle, whose size is about $2.6 \times 3.2$ cm$^2$ in process nodes such as 65 nm and it is smaller in older processes, as a result of using a stepper in for photo-litographic exposures. This limitation can be overcome, either fully by the so called stitching of reticles (large area MAPS has been already demonstrated for MAPS used in X-ray detection~\cite{Wunderer:2015joi}), or partially by butting~\cite{Theuvissen:2016blg}. An example of butting is the STAR Heavy Flavor Tracker at BNL RHIC ~\cite{Star}, where chips were butted one next to another on the staves and the  readout areas of the individual MAPS units were limited to one edge only. On the other hand, a few years after the HFT at RHIC, the upgrade of the ALICE Inner Tracker System 3 (ITS3) at the LHC is going to use the stitched arrangement of the MAPS devices~\cite{Contin:2020cbu}. Both developments are targeting inner tracking or vertexing layers for Nuclear Physics (NP) experiments, where requirements on readout speed and radiation hardness are less stringent than for high energy physics experiments. 

\begin{figure}[!htb] 
\begin{center}
 \includegraphics[width=0.9\hsize]{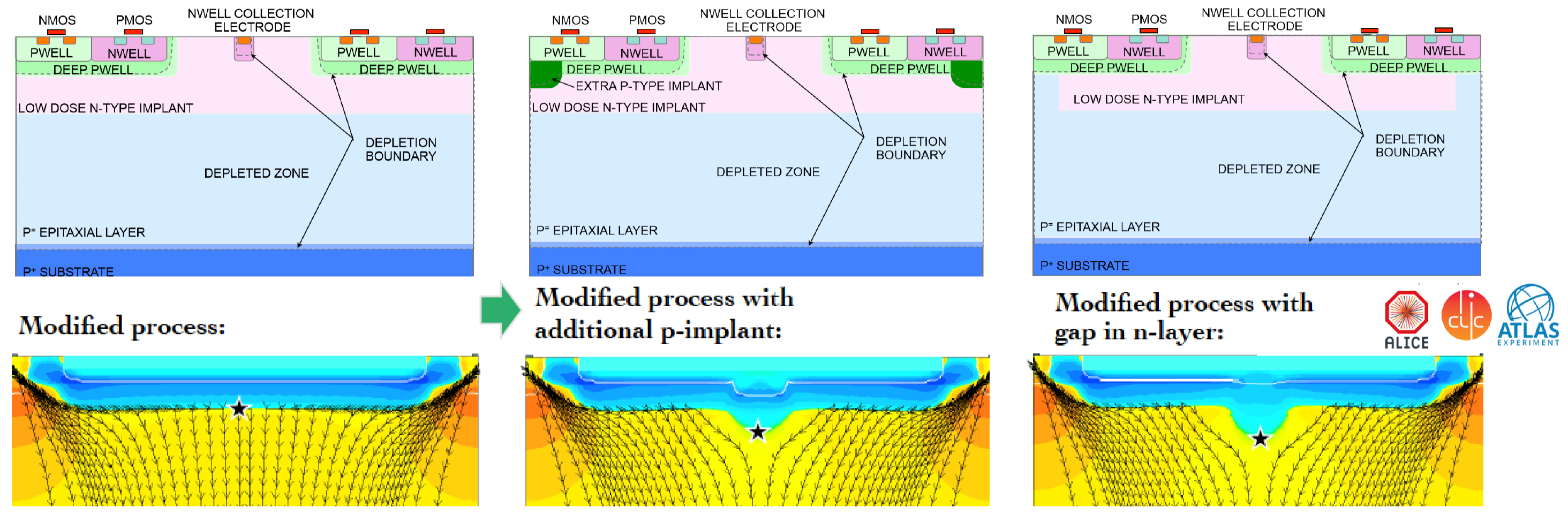}
\caption{Cross-sections of MAPS structures and electric filed lines in improved process.
\label{fig:fig_towersemi_xsection}}
 \end{center}
 \vspace{-0.7cm}
\end{figure}

Years passed since the initial demonstration of MAPS showed that full depletion of the Active Sensitive Volume (ASV) of the device is required for the charge to be collected by drift and not by diffusion, allowing a fast time response and radiation hardness. Additionally, both N and P types of transistors are required in a pixel area to develop efficient processing blocks. As a result of realizing thereof, new inner trackers, such as the ITS1/2 for ALICE, managed shifting to the increased resistivity of ASV. As an example of thereof, Figure~\ref{fig:fig_towersemi_xsection} shows cross-sections of structures built in the TowerSemi aka Tower-Jazz 180 nm. The implantations and distributions of resulting electric fields were optimized ~\cite{Snoeys:2017nim,Schioppa:2019vci}. A reasonably thick, of increased resistivity in the TowerSemi process Si film, translates to the maximally depleted ASV. Shielding of wells, visible in Figure~\ref{fig:fig_towersemi_xsection}, does not hold back the pixel electronics to only one type of transistors. The key benefit of that in tracking or vertexing, is the higher spatial resolution. The TowerSemi 180 nm process has been available for the MAPS design for a while. Very recently, an access to the TowerSemi 65 nm, process opened up, allowing its exploration to address needs for future applications. The TowerSemi 65 nm process, as opposed to the older 180 nm, allows more than four-fold increase of the number of transistors per pixel. Building a new generation of MAPS, thinned and on an increased-resistivity substrate, is an opportunity that has been enabled thanks to the developments for ITS3 of the ALICE experiment~\cite{Contin:2020cbu,Mager:2020trt}. 

Next MAPS developments will target a diverse set of goals and applications. This paper directly addresses the issues of making large surfaces by developing the intermediate stage of tens of m$^{2}$, including powering and readout of large numbers of MAPS.  The ongoing efforts will enable detectors with far more pixels per unit area, and thus higher resolution, with substantially lower material budget and significantly less expensive. The next generation of MAPS, will have to address improvements to speed and resolution performance, and the system approaches needed for large scale use at a reasonable cost. In this white paper we focus on the following applications: 
\begin{itemize}
    \item MAPS with characteristics suitable for trackers and electromagnetic calorimeters (ECal) at future colliders experiments. We will use the requirements of the SiD detector for ILC~\cite{White:2015dxj} as a benchmark, with an outlook to a broader range of the future applications. 
    \item The ongoing efforts towards dedicated MAPS for the Electron-Ion Collider (EIC) at BNL, for which the EIC Silicon Consortium was already instantiated. This aims at the High-Performance Tracking (HPT) MAPS~\cite{Collu:2020eic}. This effort is synergistic with the ALICE ITS3 developments as the performance requirements are similar.
    \item Space-born applications for MeV $\gamma$-ray experiments with MAPS based trackers (AstroPix).
\end{itemize}


\section{MAPS developments for future linear e$^+$e$^-$ collider}

\subsection{Large-area MAPS}

The detectors at future lepton machines must enable unprecedented precision measurements relating to the Higgs boson and the top quark, and search for new phenomena that could be missed at the LHC, such as electroweak particles expected in some beyond the standard models. These ambitious physics goals translate into very challenging demands on systematic errors in tracking, heavy-quark tagging, jet energy measurement, and beam polarization measurement for all these machines. High precision and low mass trackers and highly granular calorimeters, optimised for particle flow analysis, will be essential for the success of the physics program. The low duty cycle of collisions (as, for instance, in the linear e$^+$e$^-$ machines) would enable power pulsing as a desirable feature to significantly reduce the heat load, make gas cooling appropriate, and thus significantly reduce the dead material in the detectors.

The CERN WP1.2 collaboration is investigating the possibility to realize wafer-scale MAPS devices on the TowerSemi 65 nm process with the ALICE ITS3 upgrade as the main driver of this R\&D effort. Within this collaboration, SLAC is investigating challenges of wafer-scale designs optimized for detectors at linear machines, focusing in particular on Si Tracker and ECal. This effort will help identify the risks that wafer-scale MAPS pose at system-level, such as yield, power distribution and fill factor, as well as evaluate essential aspects of the integration of such devices into a detector system: \textit{i.e.}, cooling, assembly procedures, wafer thinning and handling and power delivery. 

A large part of the international research efforts on MAPS are focused on the development of detectors for circular machines, which typically require a continuous-time pixel front-end architecture~\cite{Contin:2020cbu,Mager:2020trt}. The timing performance of such architectures is currently not compatible with the requirements of low duty cycle lepton (linear) colliders, which operate with bunch spacing of a few ns.
SLAC is developing a readout circuitry optimized for such machines, leveraging the beam time structure. In particular, two techniques will be implemented. The readout electronics will adopt a power pulsing scheme: the analog front-end circuitry will be powered off during the dead-time between different bunch trains. With low duty cycle machines like \CCC and ILC, this technique enables a power reduction by more than two orders of magnitude. Power pulsing techniques were previously developed and characterised with the KPiX ASIC~\cite{kpix}. Second, the pixel front-end circuitry will be based on a synchronous readout architecture, where the operation of the circuitry is timed with the accelerator bunch train. In this way, the noise and timing performance of the circuitry can be maximized while maintaining low-power consumption. SLAC will leverage a decade of expertise with synchronous readout architectures operating with fast integration times~\cite{dragone_epix}, which have been implemented in all ASICs developed for the Linear Coherent Light Source (LCLS).

By combining all these techniques, the goal of the current R\&D at SLAC is to achieve the specifications described in in Table~\ref{tab:pixel_specs}. We have derived the initial specifications from the \CCC~configuration~\cite{ccc}, as it is the one that provides the most challenging technological case for timing resolution and is compatible with the current limits of MAPS technology. Moreover, a sparse read-out mechanism, based on asynchronous read-out logic, will minimise the digital power as well as make the circuitry more robust to local variations of the transistor performance. A first, small-scale prototype of such device is expected in late 2022.

\begin{table}[tb]
\centering
\begin{tabular}{ll}
    \hline
    \textbf{Parameter}        & \textbf{Value}           \\
    \hline
    Min. Threshold            & 140 e$^-$          \\
    Spatial resolution        & 7 \micron            \\
    Pixel size                & 25 x 100 $\mu$m$^2$    \\
    Chip size                 & 10 x 10 cm$^2$     \\
    Chip thickness            & 300 \micron          \\
    Timing resolution (pixel) & $\sim$ns       \\
    Total Ionizing Dose       & 100 kRads       \\
    Hit density / train       & 1000 hits / cm$^2$ \\
    Hits spatial distribution & Clusters         \\
    Power density             & 20 mW / cm$^2$     \\
    \hline
    \end{tabular}
    \caption{Target specifications for 65 nm prototype.}
    \label{tab:pixel_specs}
\end{table}

The development of wafer-scale MAPS will allow designers to investigate the following challenges:

\begin{itemize}
    \item \textbf{Power pulsing:} to take full advantage of the power pulsing technique, the current drawn from the supply needs to reach the peak value in the shortest time possible, minimizing the duty cycle and thus decreasing the average power consumption. However, the instantaneous current consumption of the pixel matrix can reach several Amperes over a few microseconds.
    \item \textbf{Power distribution:} the distribution of the power supply over a large area is challenging because of the non-negligible voltage drop over long metal distribution lines. 
    \item \textbf{Yield:} the probability of fabrication defects scales with the area of the device. For reticle-size MAPS, a defect in one reticle would result in a lower number of usable dies per wafer. However, a defect on a wafer-scale device is almost inevitable and could result in the loss of a full wafer. Therefore, it is essential to develop new techniques to mitigate the effects of fabrication defects, such as shorts between supply and ground lines. 
    \item \textbf{Stitching techniques:} the design of stitching MAPS  introduces additional layout design rules and methodologies, with the goal to increase the fabrication yield. This additional set of rules is not traditionally encountered by ASIC designers, therefore exposing ASIC designers to such design rules is an essential first step towards the development of wafer-scale devices.
    \item \textbf{Assembly and power delivery:} preliminary mechanical and assembly tests need to be carried out to evaluate techniques to deliver power to such sensor, while minimizing the dead material of the detector.
\end{itemize}

\subsection{MAPS Performance for Tracker detectors} 

A MAPS based tracker for SiD would feature
a sensor of size similar to that described in the
ILC TDR, 10×10 cm$^2$ devices. It would be constructed by stitching 2cm x 2cm reticles.  However, such a device would provide exceptional granularity of 25\micron~by 100\micron~pixels, with the alignment placing the 25\micron~pixel dimension in the bend direction, providing a resolution of 25\micron$/\sqrt{12} \approx 7$ \micron~without charge sharing.
The 25\micron~pixel size matches the KPiX-readout, silicon-strip width
of the SiD TDR design which was recently assembled, tested, and shown to achieve 7\micron~resolution~\cite{Lycoris}.
The depleted 10\micron~thick epi layer charge collection of the MAPS allows a minimum threshold of 1/4 MIP, ensuring high efficiency.
The pixel nature provides vastly improved pattern recognition for track finding over the strip devices.
For the endcaps, such a sensor would eliminate the need for two sensors in a small-
angle-stereo configuration, reducing both the material budget and cost.

\subsection{MAPS Performance for ECal} 

The finely granular, digital readout of the SiD ECal offered by application of MAPS sensors provides the potential for significantly enhanced performance over that envisioned in the ILC TDR~\cite{behnke2013international}.  One advantage of this digital approach over the TDR analog approach is the reduction of the effects due to variations in energy deposition, such as Landau fluctuations.  Fluctuations in the development of the shower remain as the main contribution to resolution. The fine granularity also reduces the likelihood of overlapping particles per pixel and improves the separation of nearby distinct showers, such as from high energy $\pi^0$s or jets, and contributes to improved particle flow pattern recognition. Quantifying the nature of these effects is being investigated with \geant simulations.

\begin{figure}[!htb] 
\begin{center}
 \includegraphics[width=0.8\hsize]{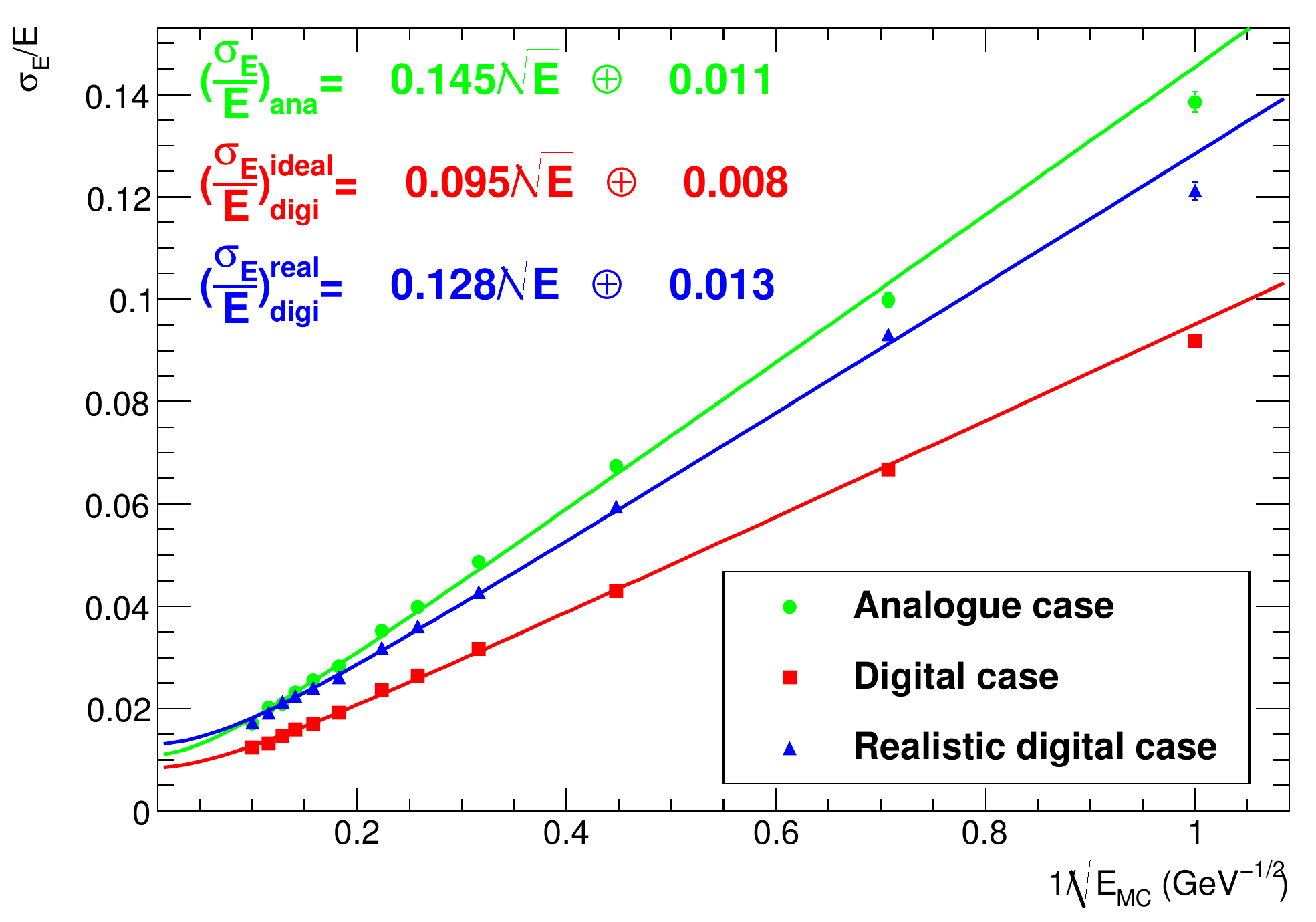}
\caption{The energy resolution as a function of the incident energy for single electrons for both analog
and digital readout using a \geant simulation. The realistic digital cases includes effects of saturation and charge
sharing, leading to a degradation of 35\%~\cite{Ballin:2009yv}.
\label{fig:fig_DECAL}}
 \end{center}
 \vspace{-0.7cm}
\end{figure}

The pixel configuration of \SI{2500}{\micron^{2}}, segmented as  \SI{25}{\micron} x \SI{100}{\micron}, is designed for the tracking performance derived from the precision of the \SI{25}{\micron} size.  Excellent performance with a purely digital ECal based on this fine granularity is expected.  
Previous studies~\cite{Ballin:2009yv,Stanitzki:2011zz,Dauncey:2010zz} have even indicated potential energy resolution advantages 
for a digital ECal solution (see Figure~\ref{fig:fig_DECAL}). 

\begin{figure}[!htb]
\begin{center}
\includegraphics[width=0.8\hsize]{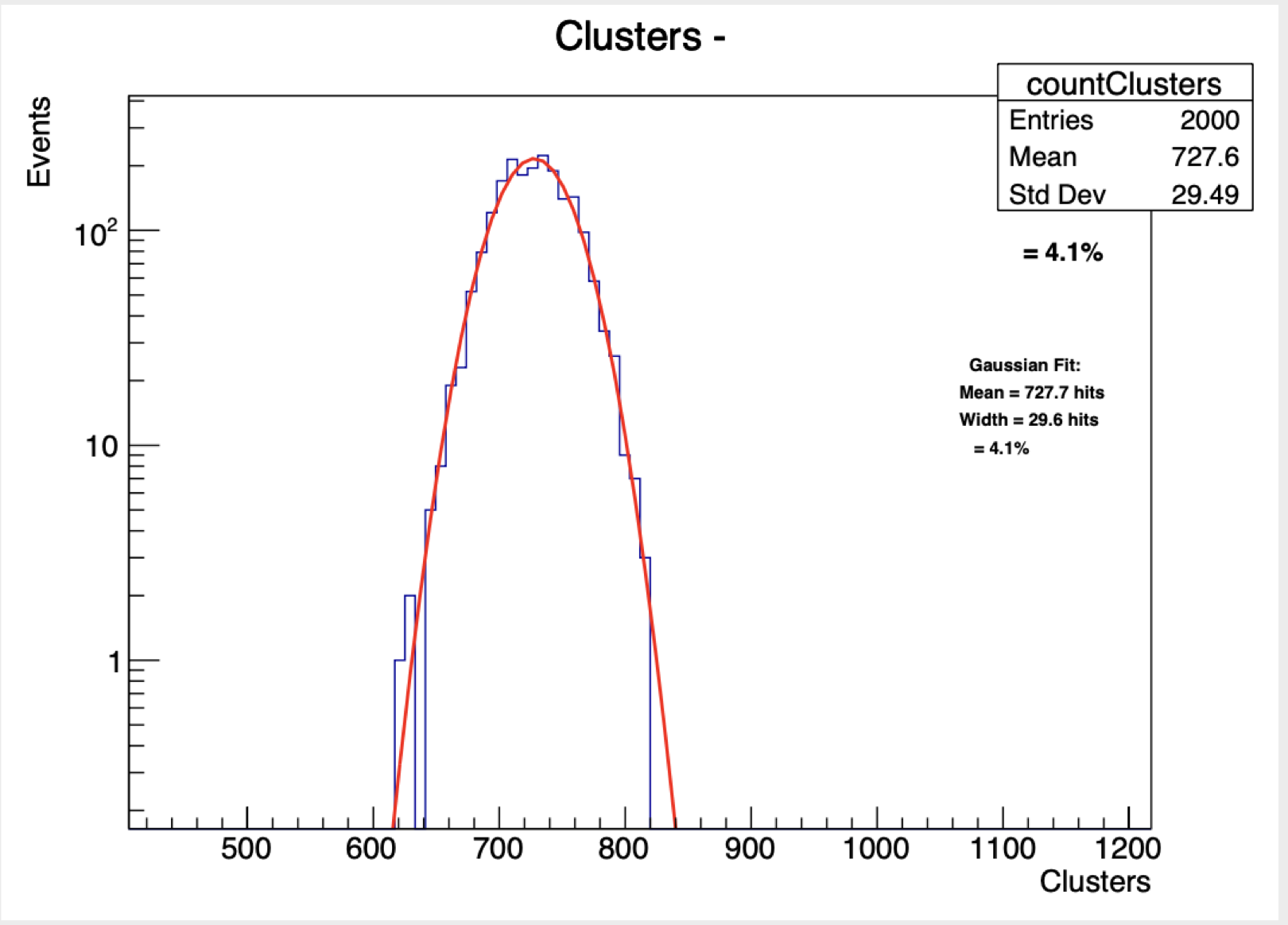}
\caption{The distribution of cluster counts for a 10~GeV electron shower in the new \SID digital MAPs design based on a \geant simulation.~\cite{Brau:2021}
\label{fig:fig_New_Eres}}
 \end{center}
 \vspace{-0.7cm}
\end{figure}

New simulation studies, based on this fine, digital configuration, have 
confirmed the previous studies referred to in Figure~\ref{fig:fig_DECAL} and 
demonstrated additional details on the performance~\cite{Brau:2021}. These 
studies indicate the electromagnetic energy resolution based on counting 
clusters of hits in the MAPS sensors should provide better performance than 
the SiD original design based on  \SI{13}{\milli\meter\tothe{2}} analog pixels, as shown in Figure~\ref{fig:fig_New_Eres}.  

\geant simulations of the performance of digital MAPS applied in the electromagnetic calorimeter has been under development and study throughout 2021, and are continuing~\cite{Brau:2021}.  These studies are aimed at understanding the ultimate performance and limitations, and to inform the ASIC designers on the requirements for the sensor chips.  The expected performance has been found to exceed the requirements and performance of the SiD TDR ECal design.  Pixel structures of 25 x 100 \micron$^2$ achieve equivalent performance to a 50 x 50 \micron$^2$ design.  The 5T magnetic field has been found to have a minor effect and to degrade the resolution by a few per cent due to the impact on the lower energy electrons and positrons in a shower.  

\begin{figure}[!htb]
\begin{center}
\includegraphics[width=0.8\hsize]{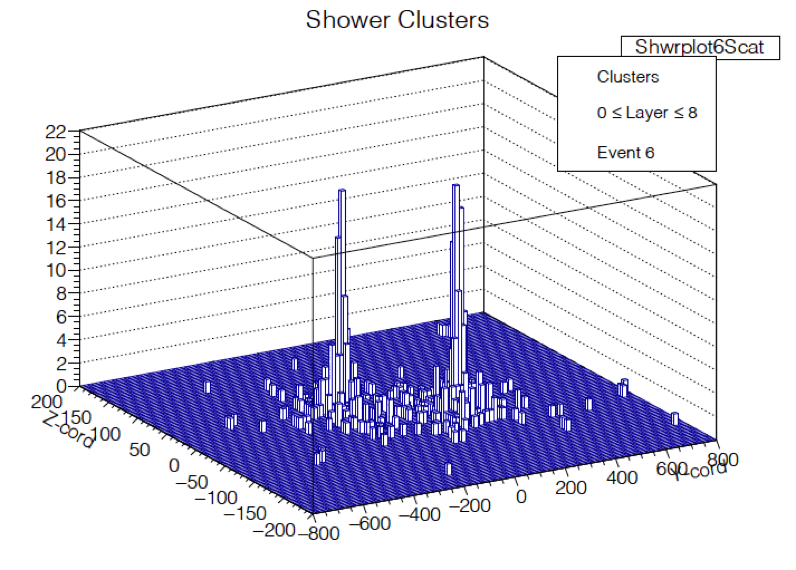}
\caption{Transverse distribution of clusters in the first 5.4 radiation lengths for two 10~GeV electron showers with a separation of one cm in the new SiD digital MAPS design based on a \geant simulation~\cite{ilcx2021}.
\label{fig:fig_New_Two}}
 \end{center}
 \vspace{-0.7cm}
\end{figure}

\begin{figure}[!htb]
\begin{center}
\includegraphics[width=0.7\hsize]{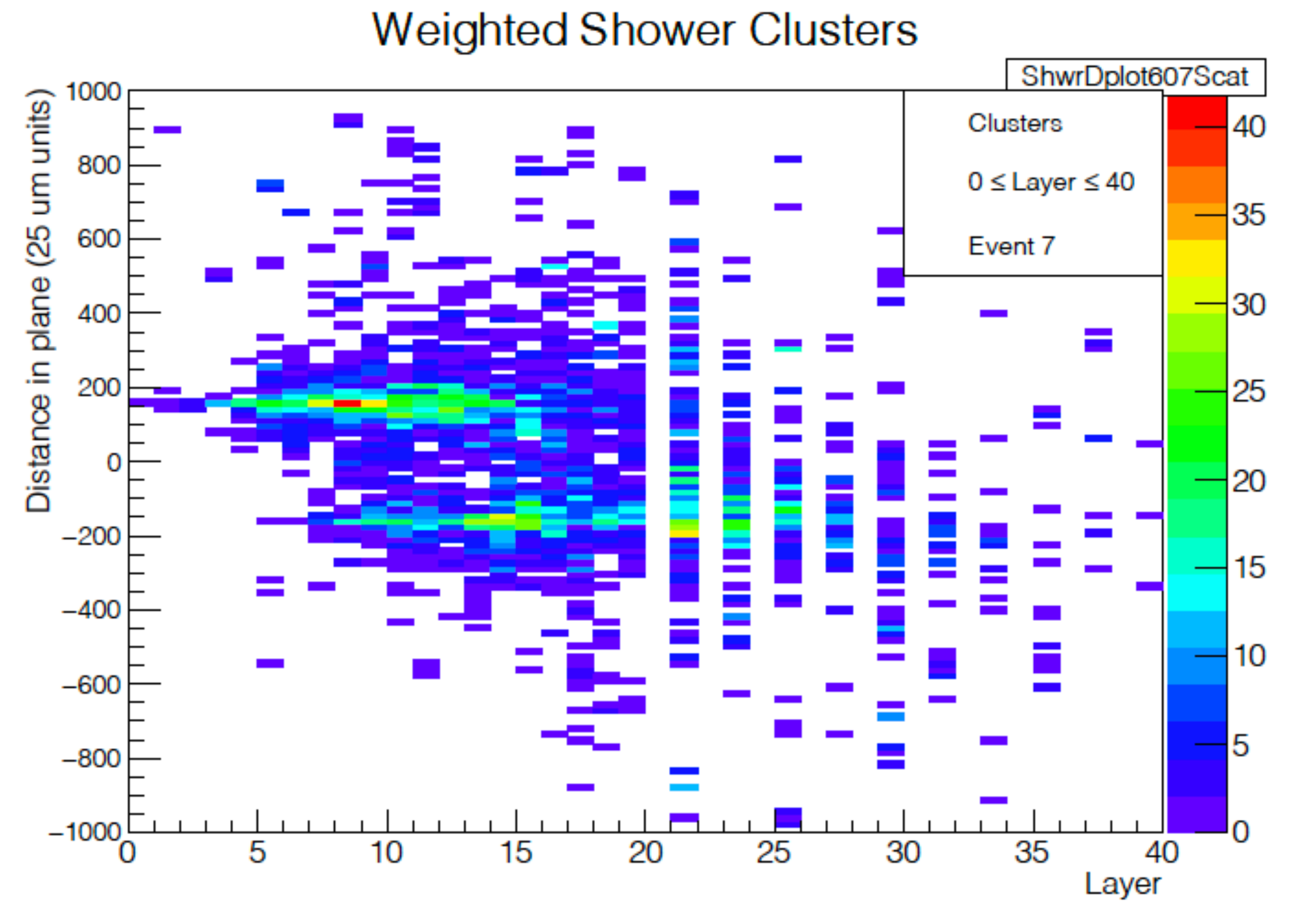}
\caption{Projection in z-layer plane of the pixel clusters in two 20 GeV gamma showers emerging from a 40 GeV $\pi^0$ decay.  The z direction is the 100 $\micron$ pixel direction and the layers shown are 20 thin (0.64 X$_0$) followed by 10 thick tungsten layers.  Each vertical bin is 400 $\micron$ wide. The two showers are separated by less than one cm~\cite{Brau:2021}.
\label{fig:fig_two_showers}}
\end{center}
\end{figure}

\begin{figure}[!htb]
\begin{center}
\includegraphics[width=0.5\hsize]{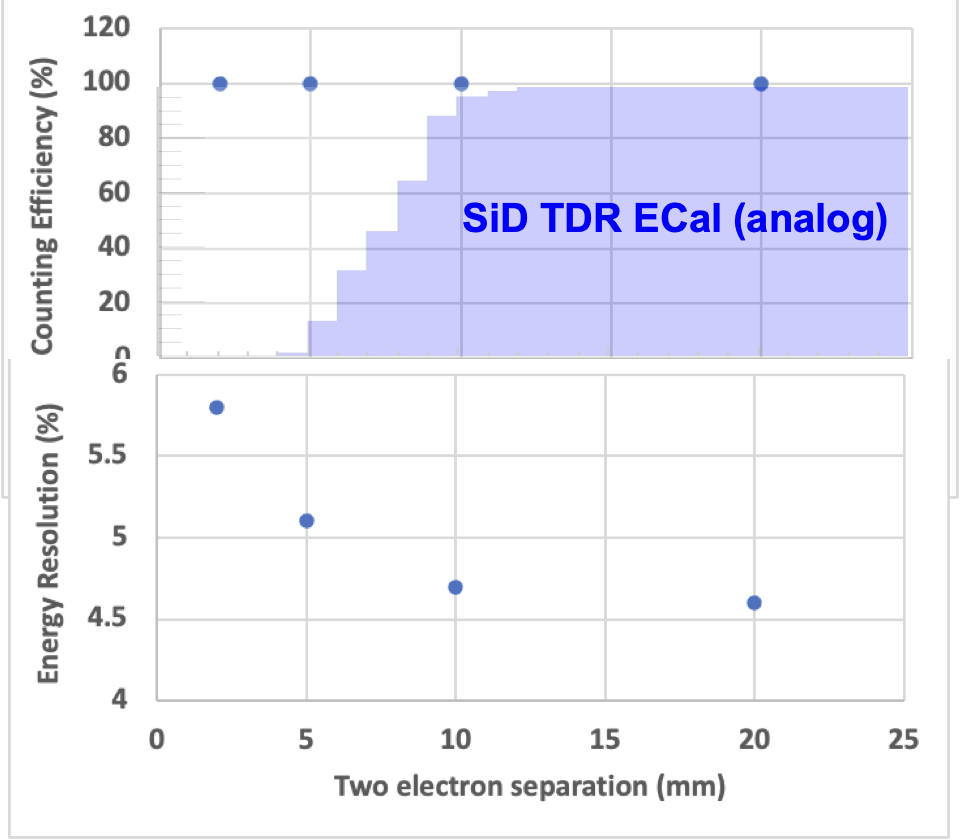}
\caption{Efficiency for distinguishing two 10 GeV electron showers as a function of shower separation (upper curve) and the degradation of energy resolution as a function of separation due to overlap of cluster hits (lower curve) in the new SiD digital MAPS design based on a \geant simulation~\cite{Brau:2021}.
\label{fig:fig_New_Sum}}
 \end{center}
 \vspace{-0.7cm}
\end{figure}

Two-showers separation is excellent, 
as shown in Figure~\ref{fig:fig_New_Two} for two 10 GeV electron showers separated by one cm and Figure~\ref{fig:fig_two_showers} of two 20 GeV gamma showers from a 40 GeV $\pi^0$ decay.  The fine granularity of pixels provides excellent separation.  The performance for two electron showers versus their separation is summarized by Figure~\ref{fig:fig_New_Sum}.  The fine granularity allows for identification of two showers down to the mm scale of separation, and the energy resolution of each of the showers does not degrade significantly for the mm scale of shower separation.

\begin{figure}[!htb]
\begin{center}
\includegraphics[width=0.7\hsize]{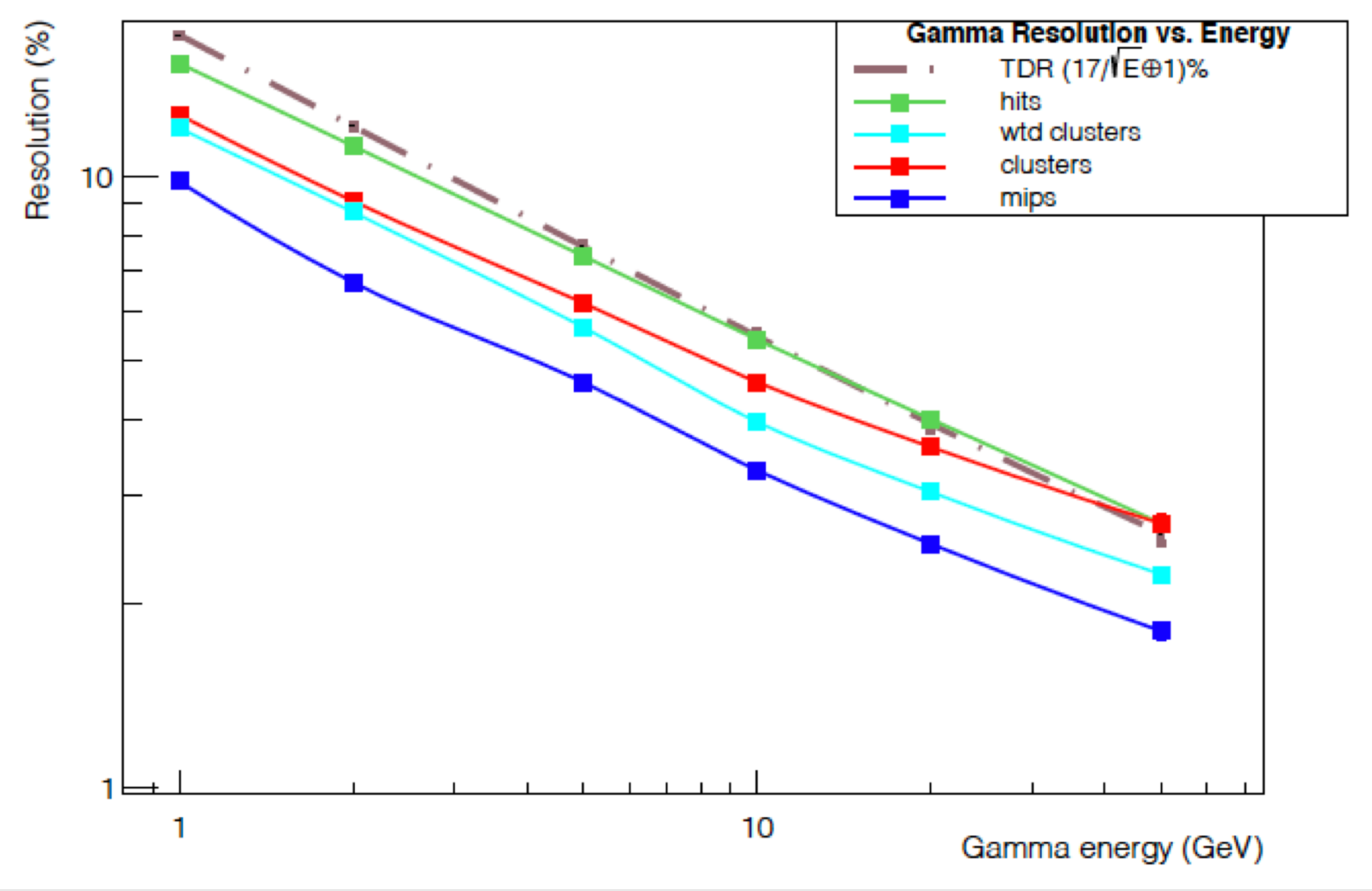}
\caption{Energy resolution for gamma showers as a function of energy.  The curves show (from lower up) the resolution based on a.) counting minimum ionizing particles (dark blue, mips), b.) modified cluster counting (light blue), c.) pure simple cluster counting (red), d.) active pixels (green, hits), and e.) the required performance from the ILC TDR (brown dash-dot)~\cite{Brau:2021}.
\label{fig:fig_resolution}}
\end{center}
\end{figure}

Figure~\ref{fig:fig_resolution} shows the gamma energy resolution performance for the range of measurements from the basic mip counting (dark blue, an idealized, best possible resolution) to that achieved by analyzing hits in clusters (light blue). These simulations are now mature and are well positioned to guide the design and production of the sensors\footnote{gamma resolution is somewhat worse than electron resolution due to the different nature of the shower development.}.
Future planned studies include the reconstruction of showers and $\pi^0$s within jets, and their impact on jet energy resolution, and the measurements of the Higgs branching ratios.

Future studies include: 
\begin{itemize}
\item Optimal Pixel size for a MAPS sensor-based ECal;
\item Comparison of analog and digital ECal performance;
\item Optimization of the overall ECal design;
\item Optimization of the design for manufacturabilty, possibly with robots.
\end{itemize}

The large volume of data provided by a MAPS based ECal reveals details of particle showers. The extraction of the most pertinent information, for example particle energy, particle type, and the separation of nearby and overlapping showers, provides an opportunity to apply Machine Learning techniques. We propose to 
apply such deep learning methods to particle and jet reconstruction in the SiD collider detector  ECal based on MAPS technology.

\section{MAPS developments for EIC}

The Electron-Ion Collider (EIC) aims at resolving the 3D partonic structure of nucleons and nuclei, to address the origin of the nucleon mass and spin, as well as the properties of a dense system of gluons. The science requirements and detector concepts for the EIC are described in~\cite{khalek2021science}.

\subsection{Performance for Tracker detectors}

MAPS technology for tracking and vertexing detectors is exploited by all the emerging detector concepts at the EIC: EIC Comprehensive Chromodynamics Experiment (ECCE), A Totally Hermetic Electron Nucleus Apparatus (ATHENA) and COmpact detectoR for the EIC (CORE). No other detector technology is capable to fulfill the requirements of compactness, high segmentation and very low material budget, which is critical to achieve the required momentum resolution. 
The EIC vertex layers follow the ALICE ITS-3 development of large-area, wafer-scale, stitched sensors that are thinned to less than 50\micron, to allow bending them around the beam pipe to achieve the required radial coverage with the expected ability of the structure to self-support, being held in place using only low-mass carbon fiber support structures ~\cite{Musa:2703140}. On the other hand the barrel tracker layers could be realized more conservatively as staves comprising smaller, but still requiring reticle stitching sensors. Regarding the disks for the tracker, the same MAPS technology as for the vertex and tracking barrel is planned.

Based on the TowerSemi 65 nm process and with reticle stitching thinning and bending of MAPS, a goal of a pixel pitch down to 10 \micron~, with power dissipation below 20 mWcm\textsuperscript{-2} to build a vertex detector, with a space point resolution of better than 5 \micron and a material thickness of 0.05\% X/X$_0$ per layer, seems to be achievable. By comparison, the current ALICE ITS (ITS2) features vertex layers with a pixel pitch of approximately 30 \micron~that dissipates 40 mWcm\textsuperscript{-2} and 0.35\% X/X$_0$ per layer. 

The EIC Silicon Consortium, of which the BNL and LBL are key members, targets co-development with ALICE-ITS3 of the wafer-scale MAPS for the vertex layers, while also developing an EIC-specific, stitched but not wafer-scale version of MAPS for the barrel layers and for the disks of the tracker. The development of the sensors is going to be assisted by the development of support structures and services.  The BNL group has been advancing developments of optimized readout architectures in order to read out data from highly granular pixel detectors. An example of such a development is the Event Driven with Access and Reset Decoder (EDWARD) architecture, based on the Loca\_Asynchronous-Global\_Synchronous topology, with the ability to respond, without built-in prioritization, to asynchronously signaled read-requests from the channels to read out data from these channels~\cite{Gorni_2022}. 

\subsection{Performance for Barrel ECal}

Physics goals at the EIC lead to unique requirements for the electromagnetic calorimeter design. In the barrel region of the electromagnetic calorimeter (barrel ECal), the momentum of scattered electrons will be measured with excellent precision ($\sigma_{p_T}/p_T (\%) = 0.1p_T \oplus 0.5 $) with the tracker. However, the electron energy and shower profile measurements play a crucial role in the separation of electrons from background pions in Deep Inelastic Scattering (DIS) processes. The ECal must also measure the energy and coordinates of neutral particles - photons, and identify single photons originating from the Deeply Virtual Compton Scattering (DVCS) process and photon pairs from $\pi^0$ decays.

The barrel region covering roughly $|\eta| < 1$, requires moderate energy resolution of approximately $\text{(10--12)}\%/\sqrt{E} \oplus \text{(1--3)}\%$, but with an excellent electron-pion separation resulting in pions suppression up to 10$^4$ at low particle-momenta below 4~GeV/$c$; a spatial resolution sufficient to separate photons from $\pi^0 \rightarrow \gamma \gamma$ decay with momentum up to about $15$\,GeV/$c$; and the capability of detecting photons with energies down to $50$\,MeV. As mentioned in the EIC Yellow Report, the required energy resolution is sufficient for the $e/\pi$ separation for particle momenta above $4\,\text{GeV}/c$ for studies of DIS processes in $e+p$ collisions at beam energies of 18×275\,GeV (the highest collision energy). The required electron-pion separation at low momenta  ($<4$~GeV/$c$) can only be achieved with the particle identification improved by using calorimeters with a much better resolution and/or by providing a shower-profile measurement capability, or by using different detectors, such as a Cherenkov detector. 

The ATHENA proto-collaboration has proposed a unique electromagnetic calorimeter that is cost effective for its excellent performance in energy and spatial reconstruction and particle identification, fulfilling the Yellow Report requirements and opening new physics opportunities~\cite{supplemental}. This section describes the proposed hybrid design concept and simulation based performance studies of the barrel electromagnetic calorimeter comprising scintillating fibers embedded in Pb and imaging calorimeter based on MAPS sensors (AstroPix).
\begin{figure}[th]
\centering
\includegraphics[width=0.6\textwidth]{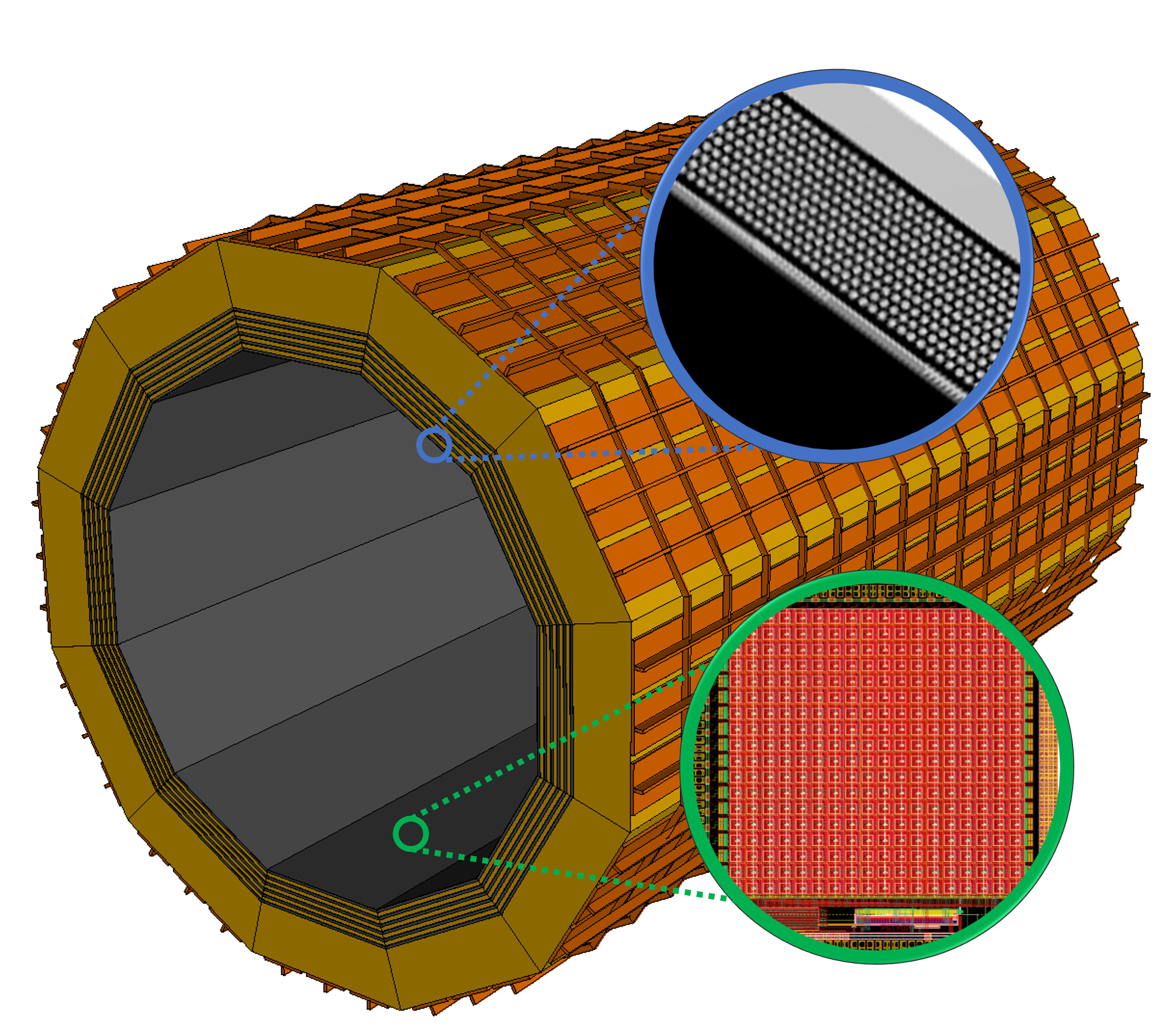}
\caption{The simulated design of barrel electromagnetic calorimeter of ATHENA. The 6 layers of imaging CMOS Si sensors are interleaved with 5 Pb/ScFi layers, followed by a large chunk of Pb/ScFi~\cite{supplemental}.}
\label{fig:barrel-ecal}
\end{figure}

\subsubsection*{Barrel ECal Concept}
The imaging layers of the barrel electromagnetic calorimeter are made up of 6 layers of MAPS sensors interleaved with 5 Pb/SciFi layers. It is followed by a large chunk of Pb/SciFi resulting in a total radiation thickness of about 20\,X$_0$. Each layer has 12 staves with the length of 405\,cm.  With the inner most imaging layer at 103\,cm and outer most one at 117\,cm, the electromagnetic calorimeter spans the area of about 172\,m$^2$ covered with CMOS sensors. The CMOS sensors are designed using a 180 nm CMOS process and developed for NASA's AMEGO-X mission concept\cite{Fleischhack:2021mhc}. A low-power CMOS sensor design is the successor of ATLASPix and called AstroPix \cite{Brewer:202109}. ATLASPiX are High Voltage MAPS, developed for the tracking applications at the LHC applications with a triggered readout~\cite{Schoning:2020zed}. The first and second versions of the AstroPix chip are being tested (details in Sec.~\ref{sec:astropixtestresults}), and a third version is under review and will be submitted for fabrication soon. These sensors have demonstrated excellent energy resolution at low energies ($\sim7\%$ at 30\,keV) coupled with low power usage and stringent cooling requirements~\cite{Brewer:202109}. The Pb/SciFi design is based on the existing GlueX barrel calorimeter~\cite{Beattie:2018xsk}. The Si imaging layers of the ECal enhance the global reconstruction performance of ATHENA by providing precision position resolution which also allows probing the 3D shower profile. The ATHENA detector with its high granularity and the ability to tag a final state radiative photon to a very small energy provides precision measurement of DIS variables \cite{arratia2021reconstructing}. The barrel ECal contains the endcap calorimeter in the electron-going direction covering the $\eta$ range of about $(-1.5, 1.1)$.

The performance goals (based on \cite{Brewer:202109}) AstroPix for pixels are listed in Table~\ref{tab:astropix_readout_specs}.

\begin{table}[h!]
\begin{center}
\begin{tabular}{l | l} 
    Pixel size & $500 \,\micron \times 500\,\micron$ \\
    Power usage & $<1$ mW/cm$^2$ \\
    Energy resolution & 10\% @ 60 keV (based on the noise floor of 5 keV) \\
    Dynamic range & $\sim 700$ keV \\
    Passive material & $<5\%$ on the active area of Si \\
    Si Thickness & $500 \,\micron$ \\
    Time tag & $\sim 1\,\mu$m\ \\
\end{tabular}
\end{center}
\caption{Performance goals for AstroPix pixels based on \cite{Brewer:202109}.}
\label{tab:astropix_readout_specs}
\end{table}

\begin{figure}[th]
\centering
\includegraphics[width=0.8\textwidth]{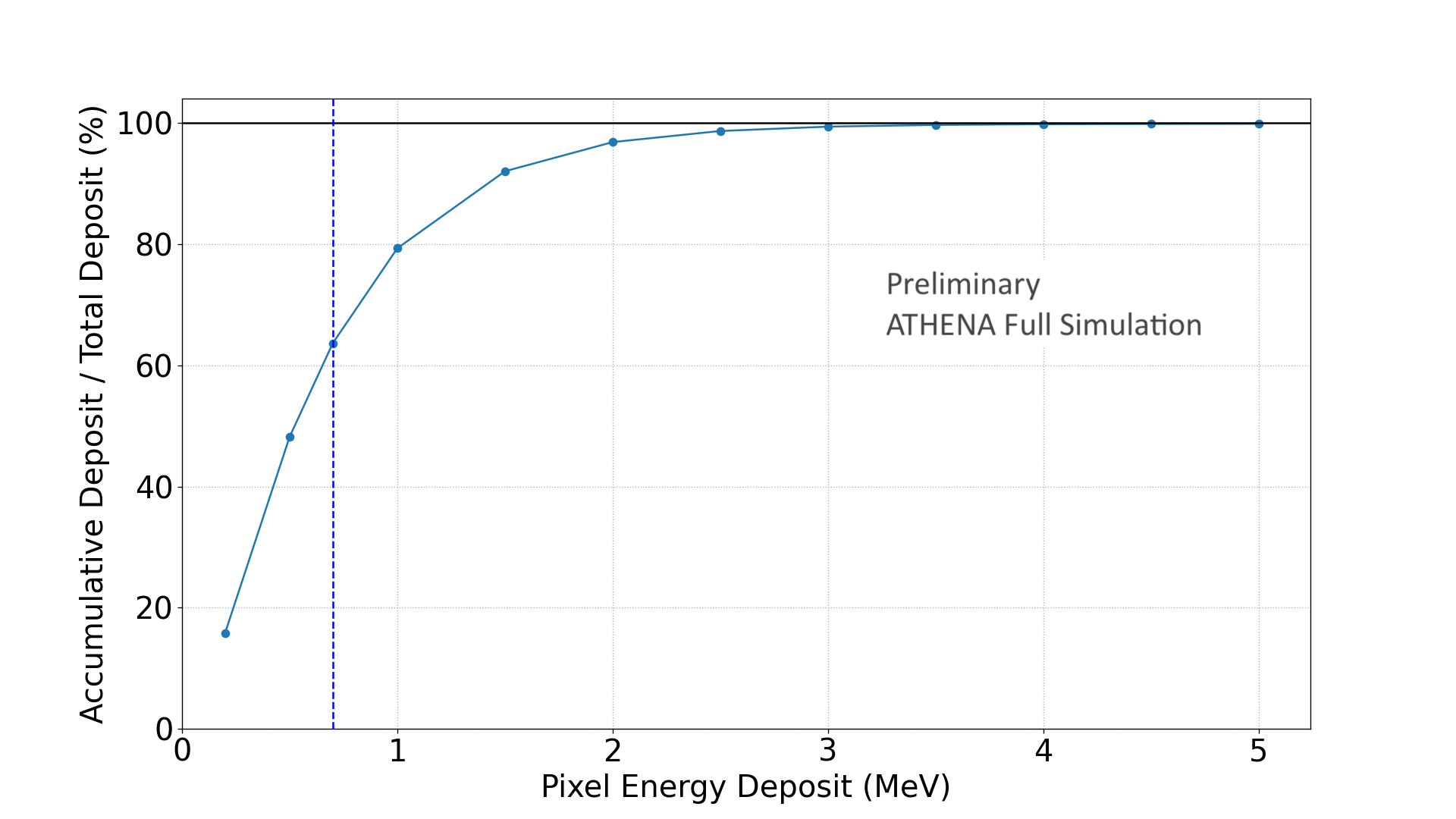}
\caption{The accumulative energy deposit over pixel energy deposit ratio for 2\,GeV particles using ATHENA simulation software. Over 99\% of the energy deposit is made through hits with a deposit $<3$\,MeV in the AstroPix pixel sensors.}
\label{fig:digitization_range}
\end{figure}

The \geant simulation studies using the ATHENA experiment software framework shows that the extension of the dynamic range of the digitized pixel readout to a few MeV could cover over 99$\%$ of the deposited energy (Fig. ~\ref{fig:digitization_range}), possibly providing better energy resolution for EIC physics. In our simulations presented in \ref{BECalPerformance}, we explore the possibility of using the AstroPix sensor off-the-shelf, with the performance parameters from Table~\ref{tab:astropix_readout_specs}. It is possible to develop a future version of the BECal pixel sensor based on the studies using the AstroPix sensor; however, as of now, we do not see a compelling need for that. 

\subsubsection*{Barrel ECal Performance}
\label{BECalPerformance}
The imaging layers of MAPS sensors provide excellent position resolution allowing precise 3D shower imaging on top of the excellent energy resolution provided by the Pb/SciFi calorimeter. The examples of the impact of the imaging layers relative to a design based only on scintillating fibers/Pb calorimetry are:
\begin{itemize}
    \item Significantly improved electron-pion separation with respect to $E/p$ method - impact on DIS cross section and asymmetries
    \item Separation of $\gamma$s from $\pi^0$ decays at high momenta up to about 40-50 GeV/$c$ and precise position reconstruction of $\gamma$s (well below 1 mm at 5 GeV) - impact on DVCS and photon physics
    \item Tagging final state radiative photons from nuclear/nucleon elastic scattering at low $x$ to benchmark QED internal corrections, by precise measurement of photon coordinates and the angle between electrons and photons
    \item Allowing PID of low energy muons that curl inside the barrel ECal (below about 1.5 GeV with 3T magnetic field) - impact on $J/\psi$ reconstruction and Timelike Compton Scattering (TCS)
    \item Improving particle identification based on other detector subsystems - providing a space coordinate for DIRC reconstruction (no need for additional large-radius tracking detector)
\end{itemize}

Performance of the barrel calorimeter is simulated using the \geant-based ATHENA simulation framework. The simulation includes realistic implementation of the detector geometry, signal digitization and readout electronics, as well as the reconstruction of the shower clusters with the barrel ECal.  

\begin{figure}[th!]
\centering
\includegraphics[width=1.\textwidth]{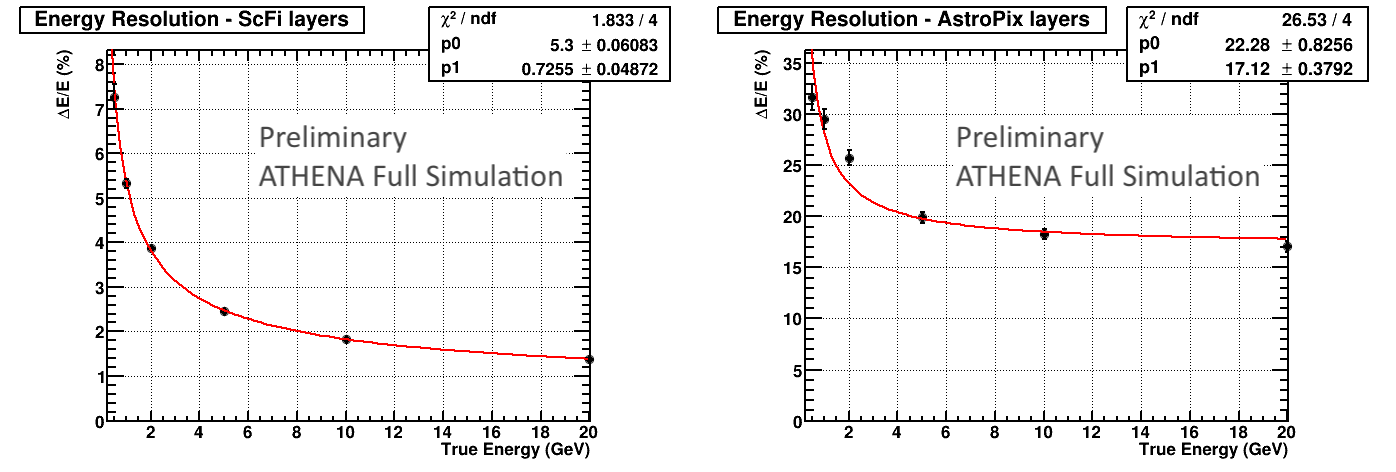}
\caption{Energy resolution for the barrel ECAL as a function of generated particle energy. The photons are generated from single-particle simulations with $\phi \in (0,2\pi)$ and $\eta = 0$. The resolution for the SciFi layers of the calorimeter is shown in left panel, and the plot on the right shows the resolution for the 6 imaging layers.}
\label{fig:energy-resolution-photons}
\end{figure}

The single-particle simulation is used to measure the energy resolution of the calorimeter with $\phi \in (0,2\pi)$ and $\eta = 0$. The shower clusters are reconstructed using imaging layers and SciFi layers and the corresponding energy resolution of photons as a function of total energy is shown in Figure~\ref{fig:energy-resolution-photons}. The energy resolution for the SciFi layers is measured to be $\sigma/E = (5.3\pm 0.1)\%/\sqrt{E} \oplus (0.73 \pm 0.05)\%$.  It is expected that the SciFi layers of the calorimeter will have significantly better energy resolution than the thin imaging Si layers, and therefore they are used for the final energy determination.   

%
The spatial resolution for the barrel ECal,is obtained using single-particle simulations, where photons were generated with energy ranging from 0.5 to 20\,GeV at normal incidence to the calorimeter layers. In global coordinates of ATHENA the particles were generated with $(\phi, \theta) = (0, \pi/2)$, which corresponds to  $(y,z) = (0,0)$. The spatial resolution of the detector depends on the granularity of the detector and energy resolution. As mentioned before, the SciFi layers provides better energy resolution while the imaging MAPS layers provide superabundant granularity with pixel size of $500\,\micron \times 500\,\micron$. The spatial resolution influences, e.g., separation of $\gamma$s from $\pi^0$ decays, precise determination of the coordinate of impact for photons, and, especially in the case of the imaging calorimetry, the $e$-$\pi$ separation. 

\begin{figure}[th!]
\centering
\includegraphics[width=1\textwidth]{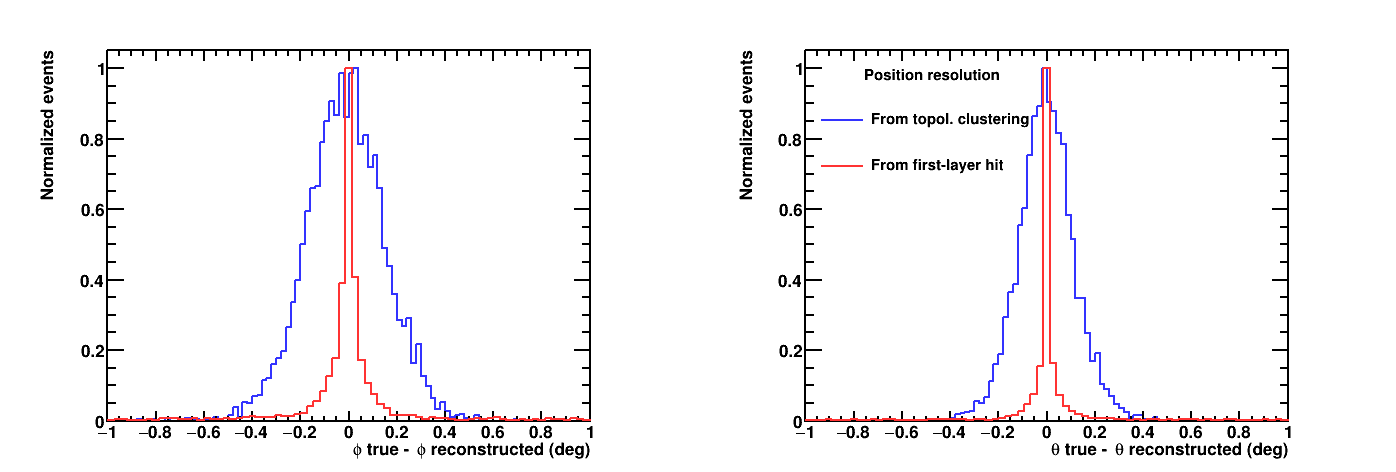}
\caption{Difference between generated and reconstructed particle $\phi$ and $\theta$ in deg using the topological clustering (blue), and obtained from the first imaging-layer hit information (red). This simulation uses 5~GeV photons generated with $(\phi, \theta) = (0, \pi/2)$ with 3T magnetic field.}
\label{fig:spatial-resolution-hit}
\end{figure}

The position resolution in $(\phi, \theta)$ or $(y,z)$ can be obtained from the reconstructed position of the cluster reconstructed with a topological cell clustering algorithm \cite{Aad2017} from the imaging layers. However, the resolution obtained with this method is strongly dependent on the chosen clustering algorithm and its optimization. A simple algorithm utilizing the information from the first imaging layer with a registered hit can already significantly improve the position resolution of the particle impact. This algorithm takes the position of the fired pixel from the first imaging layer with the registered hit. If more than one pixel fired within this layer, the position of the pixel closest to the cluster location, derived from the topological clustering algorithm, is chosen. Figure~\ref{fig:spatial-resolution-hit} presents the difference between generated and reconstructed $\phi$ and $\theta$ using the topological clustering and $\phi$ and $\theta$ obtained from the first imaging-layer hit information. Further improvements are expected from the ongoing reconstruction and clustering-algorithm optimizations, and the use of AI algorithms capable of more efficiently incorporating single-layer information and their correlation.

\begin{figure}[th!]
\centering
\includegraphics[width=.6\textwidth]{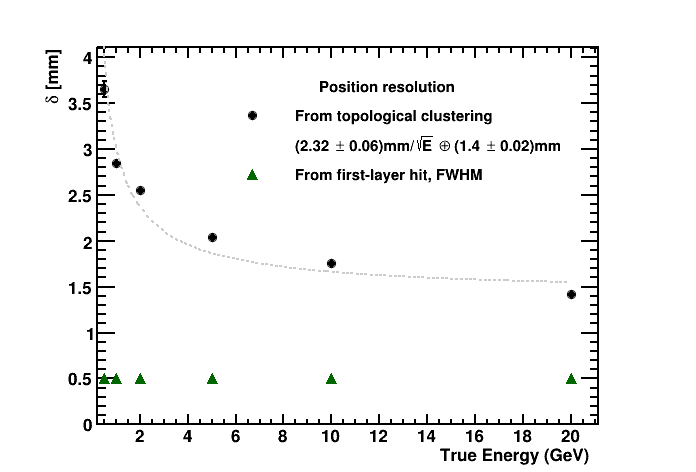}
\caption{Position resolution for the barrel ECal cluster as a function of the generated particle energy obtained from single-particle (photons) simulations of the ATHENA detector. The black dots show the resolution obtained directly from the cluster position, the green triangles show the resolution from improved algorithm using the first imaging-layer hit information determined as FWHM.}
\label{fig:spatial-resolution}
\end{figure}

The position resolution as a function of energy has been presented in Fig.~\ref{fig:spatial-resolution}. The resolution obtained directly from the cluster position is $(2.32 \pm 0.06)\,\text{mm}/\sqrt{E} \oplus (1.4 \pm 0.02) \,\text{mm}$, however, one may see that the fit which describes the spatial resolution for sampling and crystal calorimeters with large towers ($\sim$ Moli\`ere radius), doesn't describe the presented dependence ideally. The green points show the position resolution from the improved algorithm using the first imaging layer hit information determined from FWHM. One can see that for all energies the difference between true and reconstructed position at half maximum is within one $500\,\mu$m pixel.

\begin{figure}[th!]
\centering
\includegraphics[width=0.48\textwidth]{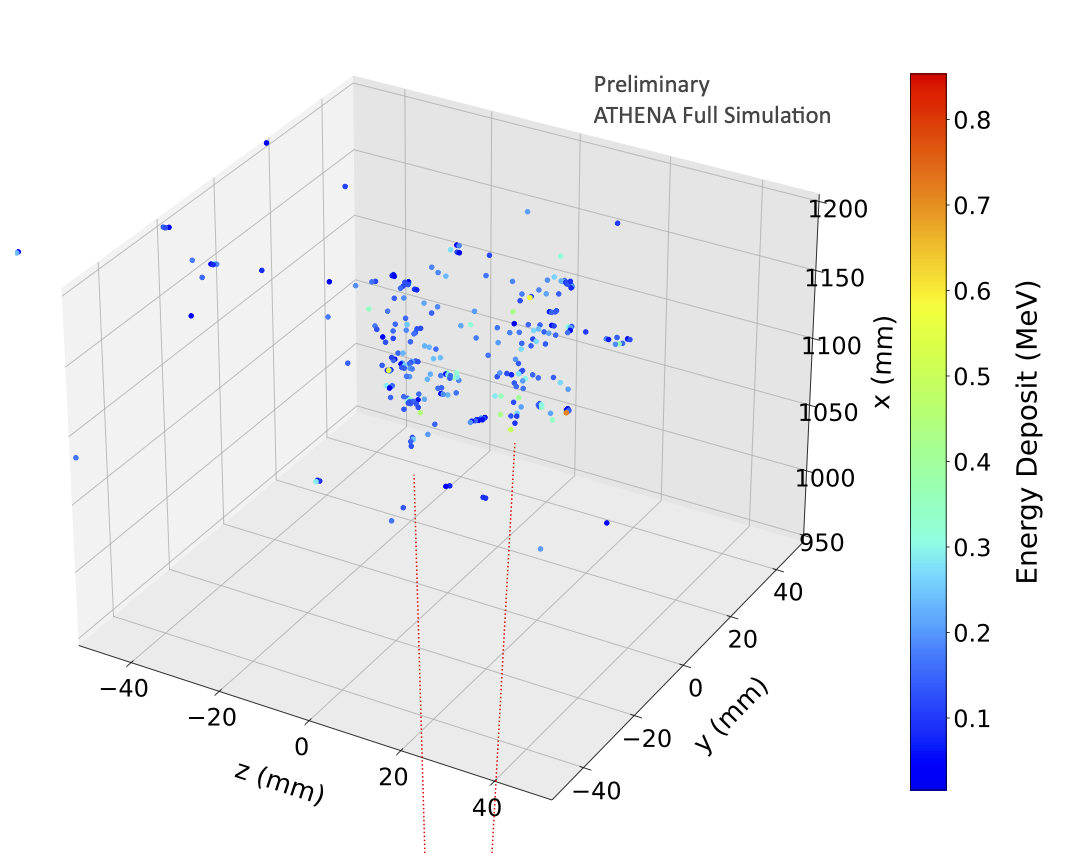}
\includegraphics[width=0.48\textwidth]{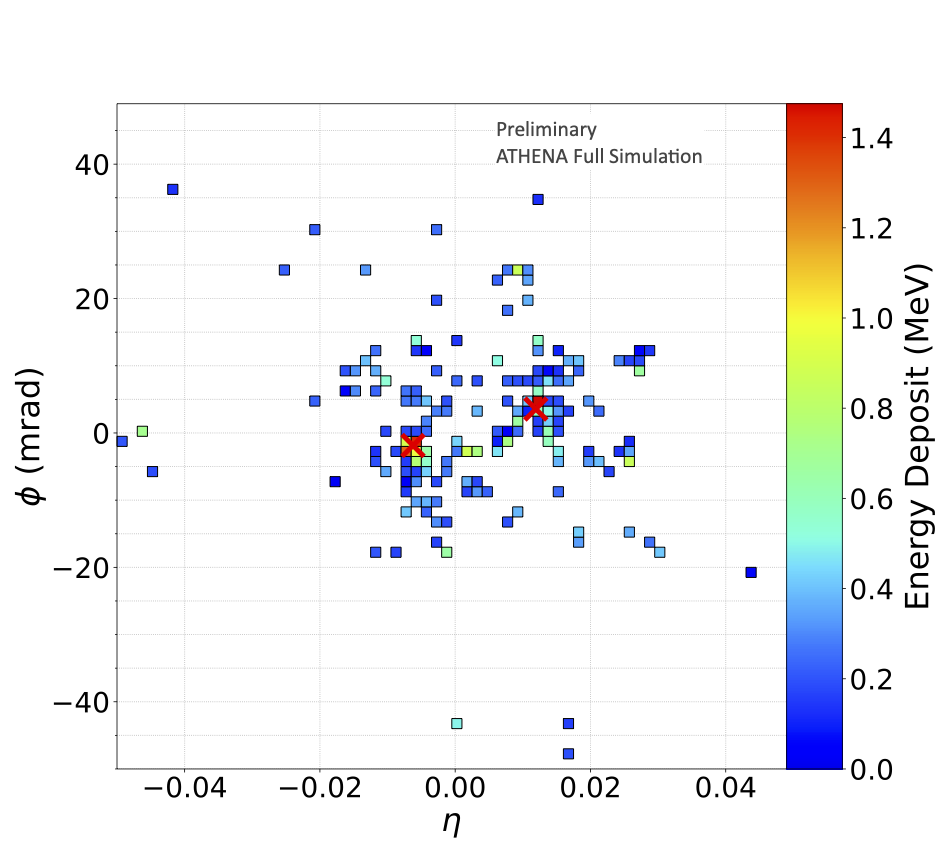}
\caption{The event display of $\pi^0 \rightarrow \gamma \gamma$ generated at 15 GeV/$c$ of pion momentum. The 3D position of hits from 6 imaging layers is shown in left panel (Red dashed lines show true $\gamma$ trajectory) . Right panel shows the cluster hits projected on the first imaging layer (True positions of $\gamma$s shown with red marker).}
\label{fig:pion0_clusters}
\end{figure}

The important characteristic of discriminating a single photon from the photon pair generated through the high momentum $\pi^0$ decay depends on the calorimeter granularity and spatial resolution. Figure~\ref{fig:pion0_clusters} shows the event display of a single $\pi^0$ at 15\,GeV/$c$ decaying to $\gamma\gamma$. The left panel depicts the 3D profile of hit positions and energy deposition in the imaging layers of the ECal. Whereas, the right panel shows the projection of the cluster hits on the first layer of the imaging calorimeter in $\eta$ and $\phi$ bins. The red cross marker describes the true position of the photons.

\begin{figure}[th!]
\centering
\includegraphics[width=0.58\textwidth]{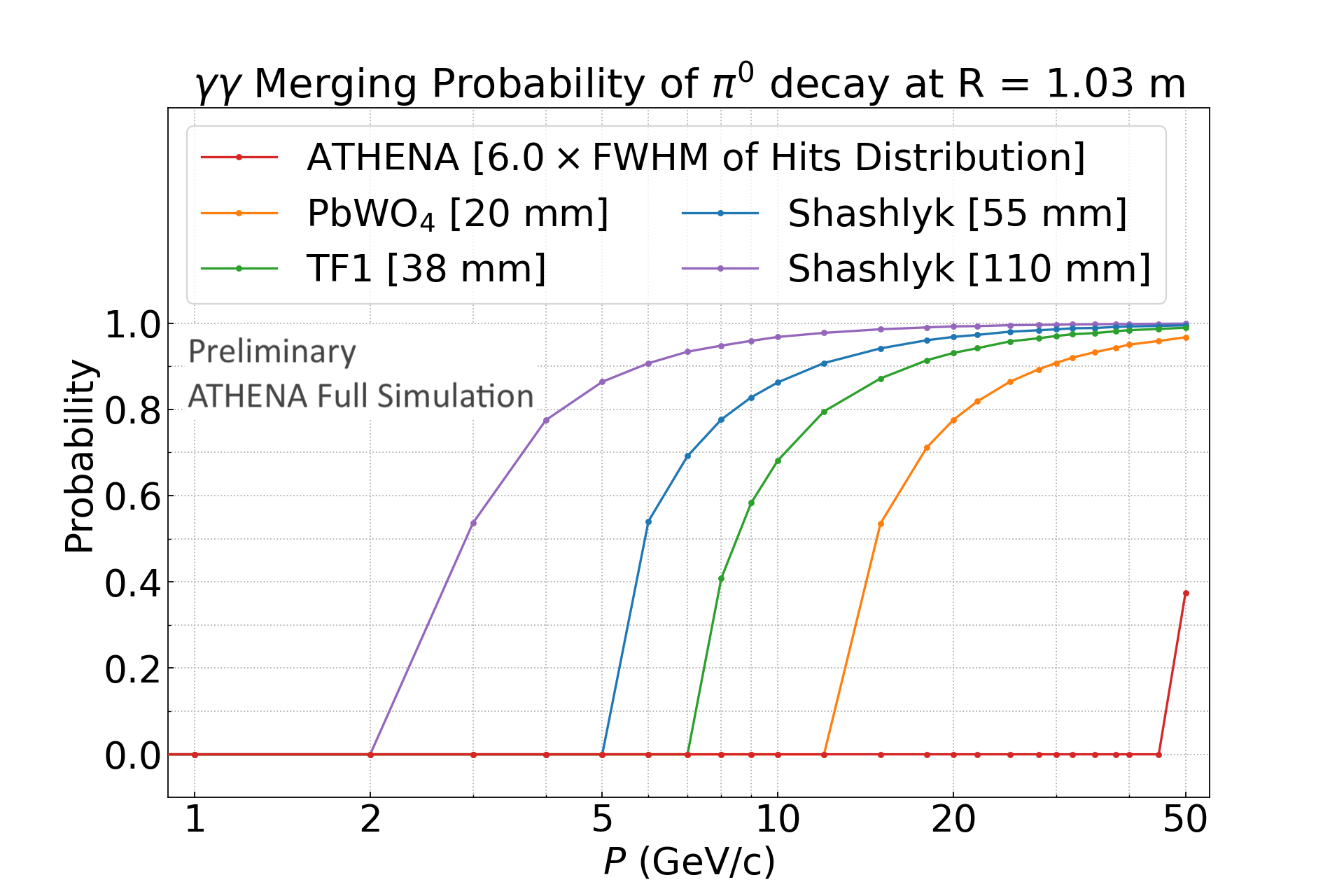}
\caption{Merging probability of the two $\gamma$s from $\pi^0$ decay in the barrel region  at $r = 103$ cm. For the ATHENA barrel ECAL the separation criteria of $6\times$FWHM of the shower profile in the $y$-coordinate.}
\label{fig:pion0-merging}
\end{figure}

Figure~\ref{fig:pion0-merging} shows the merging probability of two $\gamma$s into one cluster at $r=103$\,cm. The EIC Yellow Report~\cite{YellowReport} suggests the separation criterion of at least one cell size for the $\gamma$s detected through the ECal. For imaging ECal with the very small cell size of 500\,$\micron$, the probability of merging two $\gamma$s has been estimated using a separation criterion of $6 \times$FWHM of the shower profile at the imaging layer with the highest separation power. Fig.~\ref{fig:pion0-merging} shows that the imaging ECal allows the separation of $\gamma$s from $\pi^0$ with high momenta up to 50~GeV/$c$.

\begin{figure}[th!]
\centering
\includegraphics[width=0.4\textwidth]{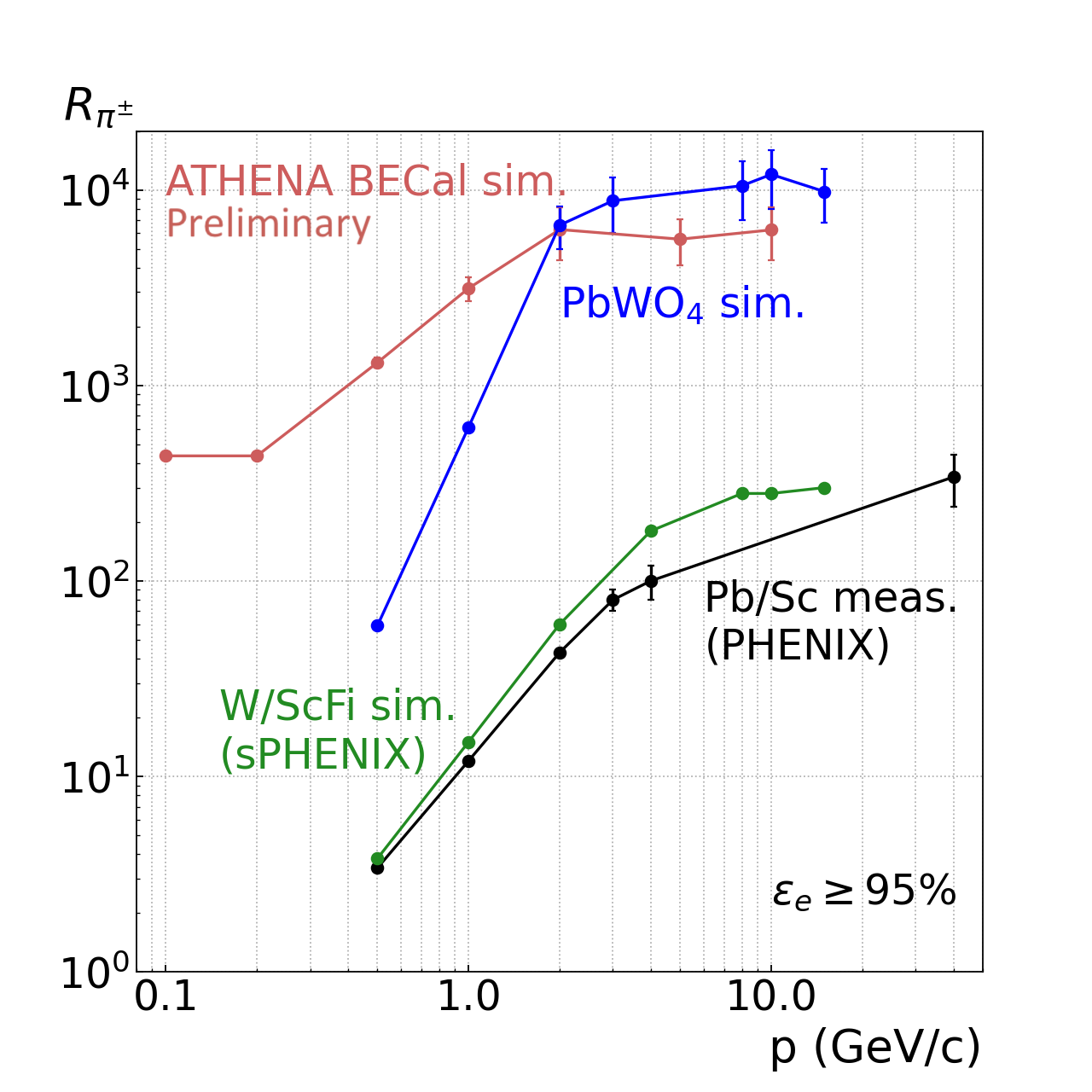}
\includegraphics[width=0.4\textwidth]{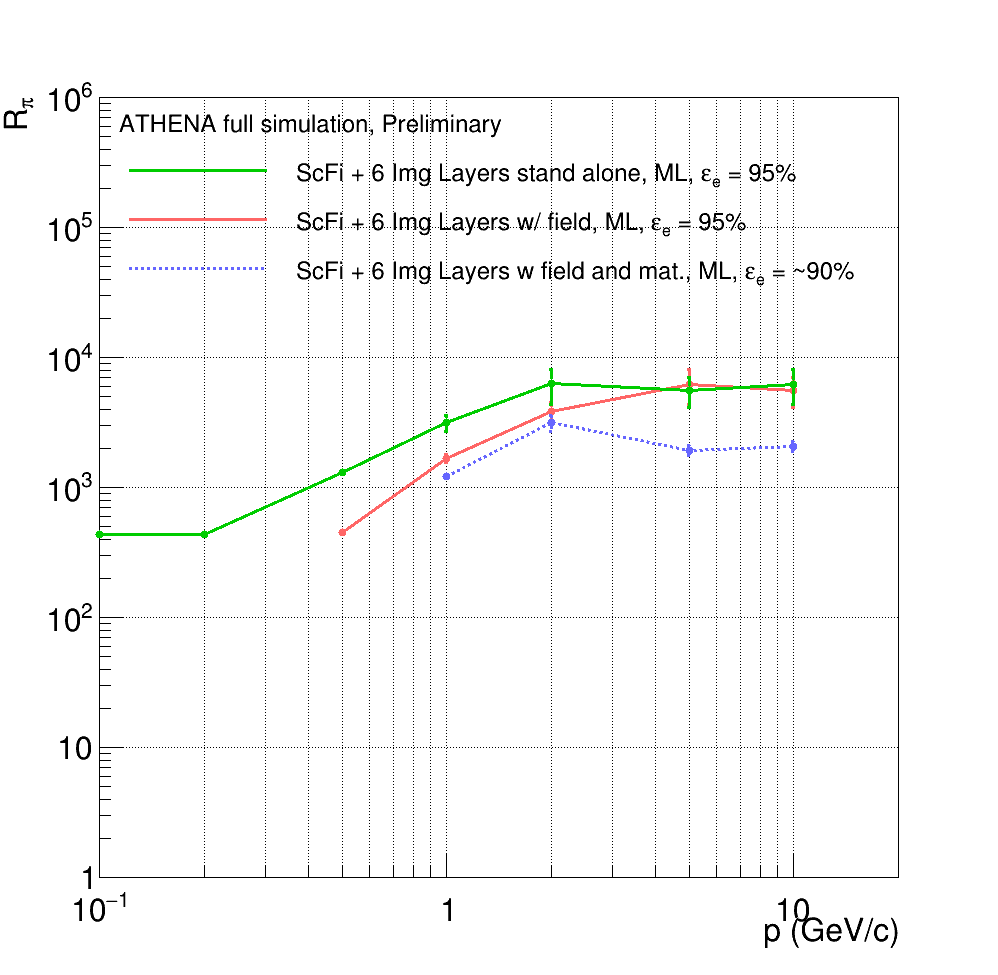}
\caption{Left: The pion rejection power of the ATHENA barrel calorimeter (red solid line) and other detectors \cite{YellowReport}. All the curves, including simulations and data, are obtained for the standalone calorimeter, \textit{i.e.}, no other materials are placed in front of the calorimeter and no magnetic field is involved. Right: The pion rejection achieved with the current status of the neural network training for the simulated standalone calorimeter without magnetic field (green line) and with the 3\,T magnetic field (red line). The blue line shows the impact of all the ATHENA material in front of the calorimeter (note that the electron efficiency drops to about 90\%).}
\label{fig:rejection_power}
\end{figure}

The proposed design of the barrel ECal aims to provide a precise electron and pion identification required for inclusive DIS physics. The barrel ECal, with its high pion rejection power in the low momentum region $p \le 2$\,GeV/$c$, outperforms the traditional calorimeter with rejection based on $E/p$ or one-dimensional shower profiles. The AstroPix layers provides a 3-dimensional profile of the electromagnetic or hadronic shower development. The overall performance of the barrel ECal obtained with a neural network classification is shown in Fig. \ref{fig:rejection_power} along with the pion rejection with $E/p$ cut for other detector technologies taken from \cite{YellowReport}. The proposed ECal detector with the AstroPix layers for 3D imaging and the machine learning algorithm for pattern recognition demonstrate a higher pion rejection power than a PbWO$_4$ calorimeter or other sampling calorimeters at $p \le 2$\,GeV/$c$, and comparable to the more expensive PbWO$_4$ at higher momentum. All measurements in the left plot of Fig. \ref{fig:rejection_power} are simulated or measured using a standalone calorimeter setup without any material in front of the ECal and no magnetic field. Including the material and 3\,T magnetic field in the simulation affects the electron efficiency. The electrons with momenta below about 0.7\,GeV/$c$ do not reach the barrel ECal. For electron energies above this value, the rejection power remains about $10^3$, with an electron efficiency drop to about 90\% (the right plot of Fig. \ref{fig:rejection_power}). 

\begin{figure}[th!]
\centering
\includegraphics[width=1\textwidth]{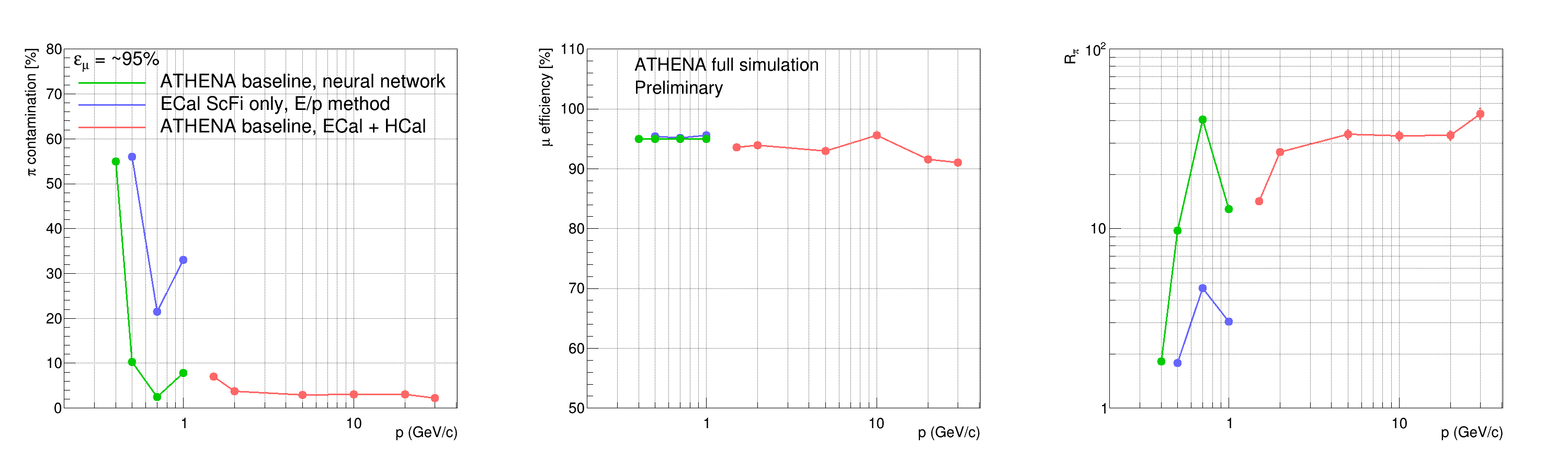}
\caption{The $\pi$-$\mu$ separation capabilities of the ATHENA barrel calorimeter. The left plot presents the pion contamination, the middle plot shows the muon efficiency, and the right plot shows the pion suppression factor. The results for the three different separation methods described in the text are presented.}
\label{fig:rejection_power_muon}
\end{figure}

The inclusion of other PID detectors is expected to improve the $e$-$\pi$ separation even further. Along with the capabilities of the electron-pion separation, the proposed barrel ECal is also capable of identifying particles with different shower profiles, like muons, with significant impact for low particle momenta (below 1.5 GeV). Fig.~\ref{fig:rejection_power_muon} shows the $\pi$-$\mu$ separation capabilities of the ATHENA barrel calorimeter. On the plot one can see two distinct momentum regions for muon PID. With 3\,T magnetic field, at $\eta=0$, muons below about 1.5 GeV/$c$ curl inside the ECal, while muons above 1.5 GeV/$c$ can reach the hadronic calorimeter (barrel HCal). For high-momentum muons, the PID criterion is a MIP-like signal required both in ECal and HCal, whereas low-energy muons are identified with a neural network classification based on information from the ScFi/Pb and imaging layers. Moreover, a separation based on an $E$/$p$ cut from one or more of the ScFi/Pb layers with the highest discrimination power has been studied. The results show that the imaging layers significantly reduce pion contamination at 95\% muon efficiency (10 times at 0.7 GeV/$c$, and 5.5 times at 0.5 GeV/$c$).

\section{Space born applications: AstroPix} \label{sec:astropixtestresults}
\begin{figure}[th]
\centering
\includegraphics[width=0.6\textwidth]{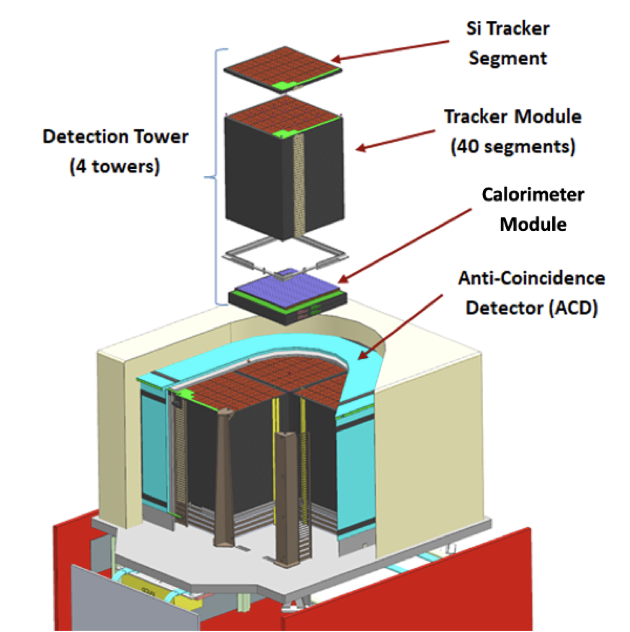}
\caption{AMEGO-X instrument design to reveal the MeV $\gamma$-ray sky with unprecedented sensitivity.}
\label{fig:amegox}
\end{figure}

The All-sky Medium Energy Gamma-ray Observatory eXplorer (AMEGO-X) is a Medium Explorer concept~\cite{Fleischhack:2021hJ}. Recent detection of gravitational wave signals and neutrinos from gamma-ray sources highlights the importance of gamma-ray observations in the multimessenger astrophysics. AMEGO-X operates both as a Compton and pair-production telescope to achieve unprecedented sensitivity between 100 keV and 1 GeV. AMEGO-X  provides better understanding of multi-messenger science and time-domain gamma-ray astronomy studying e.g. high-redshift blazars, which are probable sources of astrophysical neutrinos, and gamma-ray bursts. AMEGO-X is composed of two detector subsystems, the Gamma-Ray Detector (GRD) and the Anti-Coincidence Detector (ACD) (Figure~\ref{fig:amegox}). The GRD consists of a Tracker comprised of four square towers, each 40 cm wide and 60 cm tall with 40 layers of MAPS detectors, and a calorimeter, with four layers of Cesium Iodide (CsI) scintillator bars. The goal is to provide low power sensors to provide optimized geometry for the space applications like gamma-ray astronomy.

\begin{figure}[th!]
\centering
\includegraphics[width=0.42\textwidth]{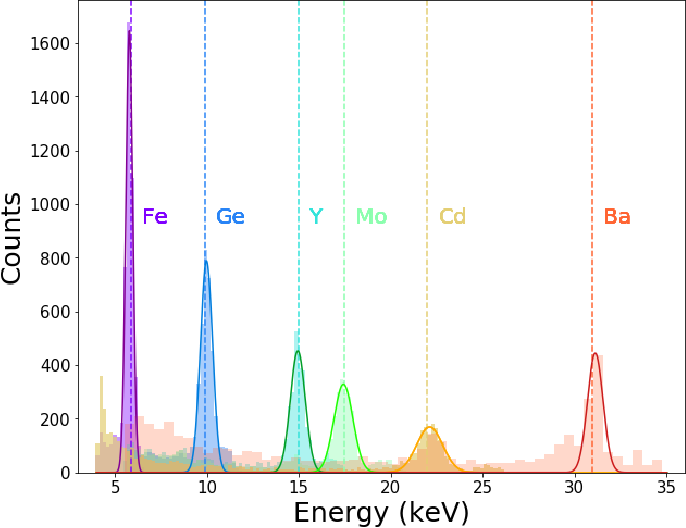}
\hspace{0.7cm}
\includegraphics[width=0.42\textwidth]{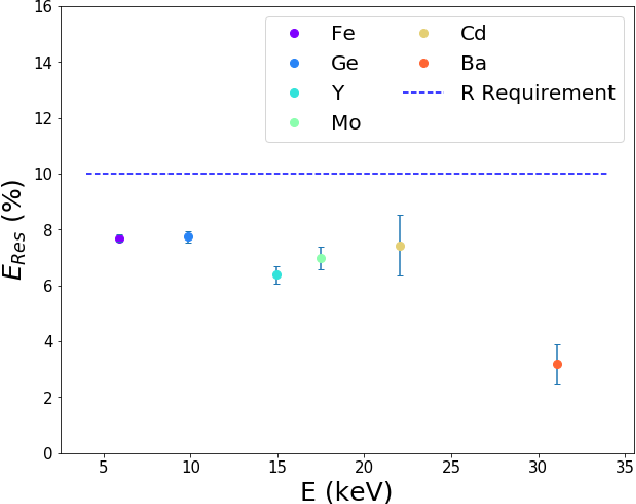}
\caption{The calibrated photopeaks represent the six different sources (left) and the analog energy resolution of the ATLASPix3 detector using a single pixel (right) ~\cite{Brewer:202109}.}
\label{fig:atlaspix3_energyresolution}
\end{figure}

As mentioned before AstroPix is a monolithic silicon pixel detector derived from ATLASPix which was developed as the pixel tracking detectors for LHC applications with a triggered readout~\cite{Schoning:2020zed}. Several prototypes and design variants were characterised in the lab and evaluated with test beam data~\cite{PERIC201999}. A latest version of ATLASPix family~\cite{tsi_astropix3}, ATLASPix3 was produced in the 180 nm CMOS process by TSI~\cite{tsi_astropix} and has a sensor area of 20.2 $\times$ 21 mm$^2$ with pixel size of 150 $\times$ 50 $\mu$m$^2$. The ATLASPix3 is fully operational and provides a timing resolution of about 10 ns \cite{Schoning:2020zed}. The ATLASPix3 sensor is evaluated to determine the potential of pixelated silicon in future space-based gamma-ray experiments. In the space applications it is critical to reduce noise in order to reconstruct low energy cosmic events ($\le$ 100 keV) with better energy resolution ($\le 10\%$ at 60 keV). The ATLASPix sensor is reconfigured for slower time response and modified for thicker sensor bulk size to absorb scattered electrons. The energy resolution and detector response of ATLASPix3 is determined using radioactive photon sources. The analog output of the ATLASPix3 sensor demonstrated the single pixel energy resolution of 7.7 $\pm$ 0.01$\%$ at 5.89 keV and 3.18 $\pm$ 0.73$\%$ at 30.1 KeV. Figure~\ref{fig:atlaspix3_energyresolution} shows the photon energy spectra (left) and related energy resolution (right) of the ATLASPix3 measured using analog output. On the other hand, the digital energy resolution of ATLASPix requires improvement as digital output is designed for MIPs and not photons. 

The AstroPix sensor is an optimized version of ATLASPix to achieve required intrinsic energy resolution of $\le 10 \%$ at 60 keV for the Compton event reconstruction in gamma-ray astrophysics. AstroPix sensors are designed with large pixel size compared to ATLASPix which results in low power consumption, as it is dominated by the amplifiers within the pixel electronics. The amplifier on each pixel is followed by low pass filter which reduces the noise. This change in circuitry reduces timing resolution but also provides required energy resolution. The current prototype of AstroPix sensors, AstroPix v2 is designed with amplifier and comparator for each pixel all within the sensitive area. AstroPix v2 has a pixel size of 250 $\times$ 250 $\micron^2$. AstroPix v2 is comprised of an 35 by 35 pixel matrix where the chip are of 10 $\times$ 10 mm$^2$. The analog output can be read from the first row through a daisy chain. The power consumption of the AstroPix v2 is currently being tested. The design review of AstroPix v3 is in progress with pixel size of 500 \micron~ with sensor area of 20 $\times$ 20 mm$^2$ and lower power dissipation. It is expected to be submitted for fabrication in April 2022.

\section{Executive Summary}

Next MAPS developments will target a diverse set of goals and applications and will have to address improvements to speed and resolution performance, and the system approaches needed for large scale use at a reasonable cost.

Future colliders will need large areas of silicon sensors, several hundred m$^{2}$, for low mass trackers and sampling calorimetry~\cite{behnke2013international}. Trackers will require multiple layers, large radii, and micron scale resolution. Sampling calorimeters will also have very large areas and are improved by very thin overall packages, which reduces the Moliere radius. MAPS with characteristics suitable for trackers and electromagnetic calorimeters at future colliders experiments are being developed. There is an on going effort that focuses on developing readout electronics compatible with a power pulsing scheme: the analog front-end circuitry will be powered off during the dead-time between different bunch trains. With low duty cycle machines like \CCC and ILC, this technique enables a power reduction by more than two orders of magnitude. Second, the pixel front-end circuitry will be based on a synchronous readout architecture, where the operation of the circuitry is timed with the accelerator bunch train. In this way, the noise and timing performance of the circuitry can be maximized while maintaining low-power consumption.  The development of wafer-scale MAPS will allow designers to investigate the power pulsing, power distribution, yield, stitching techniques, assembly and power delivery.

For the vertex detector at EIC the target point resolution is better than 5 \micron with a material thickness of 0.05\% X/X$_0$ per layer. A solution based on the TowerSemi 65 nm process and with reticle stitching thinning and bending of MAPS, with a goal of a pixel pitch down to 10 \micron~ and power dissipation below 20 mWcm\textsuperscript{-2} seems promising. This effort is synergistic with the ALICE ITS3 developments as the performance requirements are similar.

Space-born applications for MeV $\gamma$-ray experiments with MAPS based trackers (AstroPix). The AstroPix sensor is an optimized version of ATLASPix to achieve required intrinsic energy resolution of $\le 10 \%$ at 60 keV for the Compton event reconstruction in gamma-ray astrophysics. AstroPix sensors are designed with large pixel size compared to ATLASPix which results in low power consumption, as it is dominated by the amplifiers within the pixel electronics.

\section{Acknowledgements}

The work of Martin Breidenbach, Angelo Dragone, Norman Graf, Tim K. Nelson, Lorenzo Rota, Julie Segal, Christopher J. Kenney, Ryan Herbst, Gunther Haller, Thomas Markiewicz, Caterina Vernieri, Charles Young is supported  by Department of Energy Contract DE-AC02-76SF00515. The work of James Brau, Nikolai Sinev, David Strom is supported by Department of Energy grant DE-SC0017996. The work of Whitney Armstrong, Manoj Jadhav, Sylvester Joosten, Jihee Kim, Jessica Metcalfe, Zein-Eddine Meziani, Chao Peng, Paul E. Reimer, Marshall Scott, Maria \.{Z}urek is supported by the U.S. Department of Energy, Office of Science, Office of Nuclear Physics and Laboratory Directed Research and Development (LDRD) funding from Argonne National Laboratory, provided by the Director, Office of Science, of the U.S. Department of Energy under Contract No. DE-AC02-06CH11357. This manuscript has been authored by employees of Brookhaven Science Associates, LLC under Contract No. DE-SC0012704 with the U.S. Department of Energy.


\addcontentsline{toc}{section}{Bibliography}

\bibliographystyle{atlasnote}
\bibliography{bibliography.bib}

\end{document}